\documentclass[preprint]{revtex4}
\usepackage{epsfig}
\usepackage{multirow}

\def\be{\begin{equation}}
\def\ee{\end{equation}}
\def\bea{\begin{eqnarray}}
\def\eea{\end{eqnarray}}

\newcommand{\gev}{\mbox{\ensuremath{\mathrm{GeV}}}}
\newcommand{\mev}{\mbox{\ensuremath{\mathrm{MeV}}}}
\newcommand{\kev}{\mbox{\ensuremath{\mathrm{keV}}}}
\newcommand{\ev}{\mbox{\ensuremath{\mathrm{eV}}}}
\newcommand{\fb}{\mbox{\ensuremath{\mathrm{fb}^{-1}}}}

\newcommand{\jpsi}{\ensuremath{J/\psi}}
\newcommand{\x}{\ensuremath{X(3872)}}

\newcommand{\cc}{\ensuremath{(c\bar{c})}}
\newcommand{\gamg}{\ensuremath{{\gamma\gamma}}}

\begin{document}
\vspace*{4cm}
\title{\MakeUppercase{Exotic/charmonium hadron spectroscopy at Belle and BaBar}}

\author{\MakeUppercase{Dmitri Liventsev}, on behalf of Belle collaboration}



\affiliation{Institute for Theoretical and Experimental Physics,\\
B. Cheremushkinskaya 25, 117218 Moscow, Russia}
\begin{abstract}
A brief review of experimental results on
charmonium and charmonium-like hadron spectroscopy at $B$-factories is
presented. A special focus is put on recent results of $\eta_c$ and
$\eta_c(2S)$ study, $\x$ radiative decays, $\omega \jpsi$ final state
study and search for charmonium production in radiative $\Upsilon$
decays.
\end{abstract}

\maketitle

Conference talk presented at Rencontres de Moriond 2011, QCD and High Energy Interactions session, March 20--27 2011, La Thuile, Italy.

\section{Conventional and ``exotic'' charmonium states}
The first charmonium state $\jpsi$ was discovered in
1974~\cite{jpsi}. Then in six years nine more \cc\ states were
observed. No new states were found during next 22 years, until in 2002
Belle reported the detection of $\eta_c(2S)$~\cite{etac}. In 2003 Belle
discovered \x~\cite{x3872}, which marked the beginning of ``exotic'', or
unconventional charmonium-like states era. Such states decay in ways,
peculiar to usual charmonium, but have masses, widths, quantum numbers
and decay ratios, which can hardly be explained by the classical
quark-parton model. Since then two conventional and more than dozen
``exotic'' charmonium states were reported. Comprehensive review of
their characteristics, possible explanations etc can be found
in~\cite{rev}. In this paper we report some recent experimental results
on this topic from $B$-factories.

\section{$\eta_c$ and $\eta_c(2S)$}
Although $\eta_c$ and $\eta_c(2S)$ have been around for some time and
studied by different experiments, there is still large spread in their
mass and width measurements~\cite{pdg}. Moreover, our knowledge of
hadronic decays of these charmonia is rather poor. Both Belle and
BaBar performed recently new measurements of $\eta_c$ and $\eta_c(2S)$
characteristics.

BaBar claimed that $\gamma \gamma \to \eta_c \to K_S K^\pm \pi^\mp$ is
the ``right place'' for such study since Breit-Wigner line shape is
appropriate approximation here~\cite{1231}. With data set of $469\,\fb$
mass and width of $\eta_c$ were measured relative to \jpsi. In the same
paper transition form factor in $\gamma \gamma \to \eta_c$ decay was
measured and nice agreement with pQCD was observed. BaBar also reported
mass and width measurement of $\eta_c(2S)$ in the same production
process~\cite{1232}.

Belle took another approach. They studied $B^\pm \to K^\pm\eta_c
(\eta_c(2S))$, $\eta_c (\eta_c(2S)) \to (K_SK\pi)^0$ decay chain and
consistently took into account interference between decay under study
and nonresonant decay into the same final state~\cite{987}. Results,
obtained with and without interference are quite different, which means
that taking it into account is important.

Until recently only one decay mode of $\eta_c(2S)$ was known,
$\eta_c(2S) \to (K_SK\pi)^0$. Decays to 4-prong final state have not
been observed~\cite{eee}. Belle with 923\,\fb studied decays to 6-prong
final states: $6\pi$, $2K4\pi$, $4K2\pi$, $K_SK3\pi$~\cite{234}.
$\eta_c(2S)$, as well as $\chi_{c0}$ and $\chi_{c2}$, were clearly seen
in $6\pi$, $2K4\pi$, and $K_SK3\pi$ distributions. BaBar looked at
$K^+K^-\pi^+\pi^-\pi^0$ invariant mass spectrum from \gamg\ process and
found $\eta_c(2S)$ signal, as well as $\eta_c$, $\chi_{c0}$ and
$\chi_{c2}$~\cite{1232}.

\section{$\x$ radiative decays}
The \x\ was discovered by Belle as a narrow peak in $\jpsi\pi^+\pi^-$
invariant mass from $B^\pm \to \jpsi\pi^+\pi^-K^\pm$
decays~\cite{x3872}. It was confirmed by CDF~\cite{ggg}, D0~\cite{hhh}
and BaBar~\cite{iii}. Among newly observed ``exotic'' charmonium-like
states \x\ is the most studied one. It has very small width
$\Gamma<2.3\,\gev$ at 90\%\,CL for a state above open charm
threshold. Its mass is very close to $D^0D^{*0}$ threshold,
$M(\x)-(m_{D^0}+m_{D^{*0}})=-0.32\pm0.35\,\gev$. In decays to
$\jpsi\pi^+\pi^-$ invariant mass of $\pi\pi$ pair is consistent with
originating from $\rho \to \pi^+\pi^-$, indicating $C=+1$ parity of
\x. Since all charmonia are isospin singlets, decays to $\jpsi \rho$
violate isospin and should be strongly suppressed. CDF studied angular
distributions in $\x \to \jpsi\pi^+\pi^-$ decay and concluded that
possible $J^{PC}$ assignments for \x\ are $1^{++}$ and
$2^{-+}$~\cite{kkk}.

There are several unoccupied charmonium levels with appropriate quantum
numbers but their predicted masses are either too high ($\chi'_{c1}$,
$J^{PC}=1^{++}$) or too low ($\eta_{c2}$, $J^{PC}=2^{-+}$). The whole
set of \x\ characteristics also makes it hard to describe \x\ as a
conventional charmonium. Proximity of \x\ mass to $D^0D^{*0}$ threshold
led to a suggestion, that it may be a molecule-like $D^0D^{*0}$ bound
state~\cite{lll}.

Weighty argument in distinguishing between different possibilities are
radiative decays $\x \to \gamma \psi'$ and $\x \to \gamma \jpsi$. If
\x\ is a charmonium state $\chi'_{c1}$, partial with of $\x \to \gamma
\psi'$ decay should be larger than that of $\x \to \gamma \jpsi$ by
more than factor of ten~\cite{mmm}. In case of molecular state or
$\eta_{c2}$ the situation is reversed and $\gamma \jpsi$ mode is
favoured~\cite{nnn1,nnn2}.

\begin{figure}[htb]
\begin{tabular}{ccc}
\includegraphics[width=0.33\textwidth,clip=true]{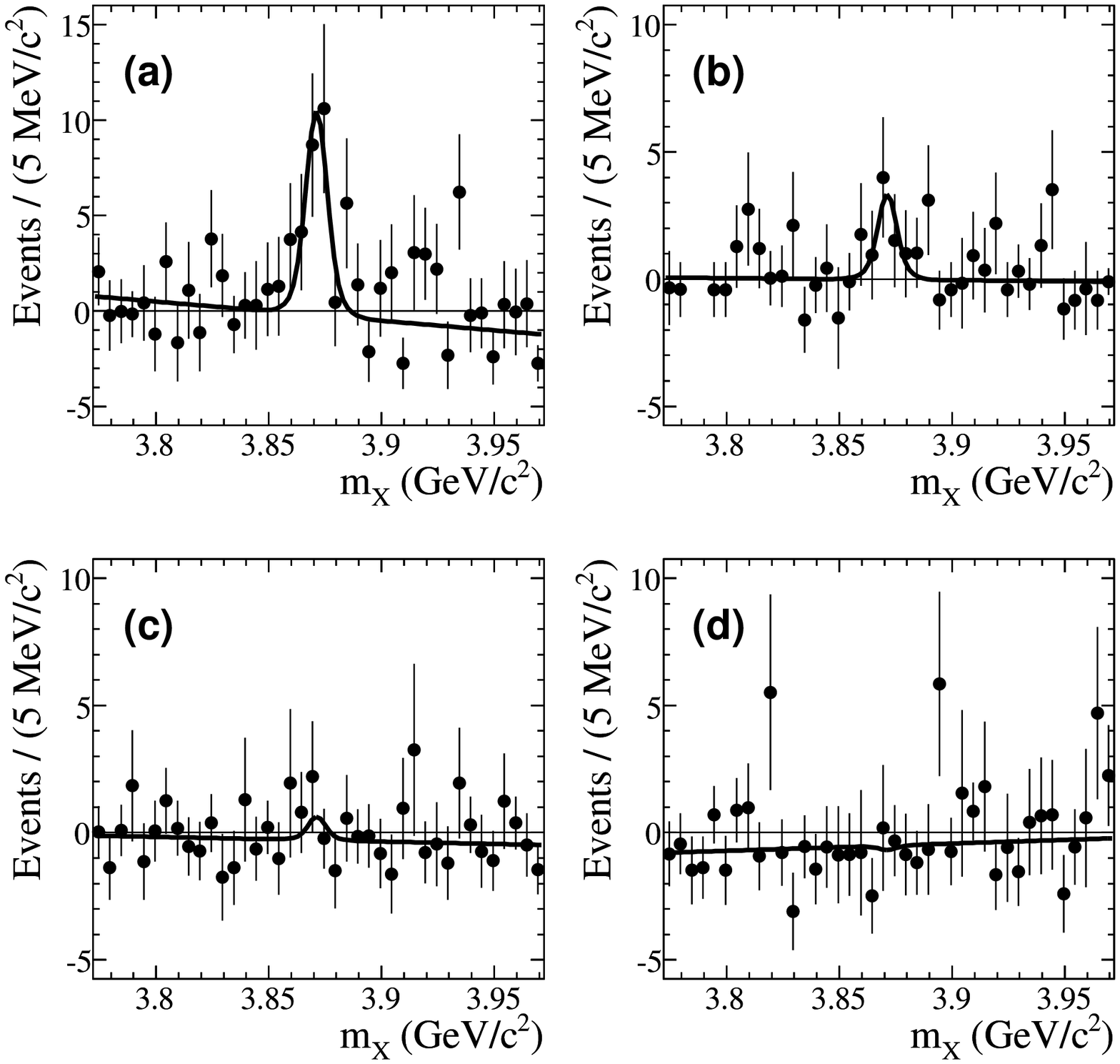}&
\multicolumn{2}{c}{\includegraphics[width=0.66\textwidth,clip=true]{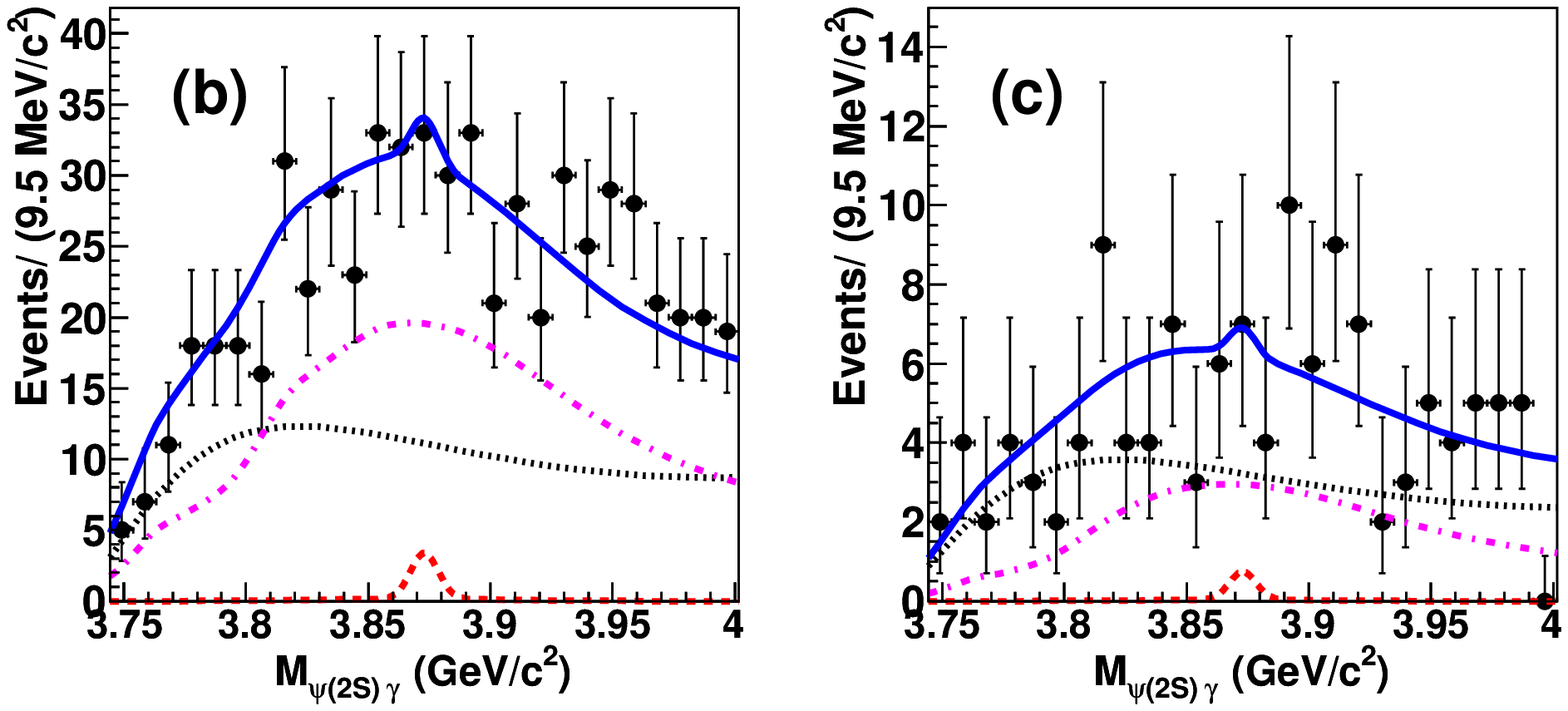}}\\
\end{tabular}
\caption{The $\gamma \psi'$ invariant mass distribution for (a) $B^+ \to
  \gamma \psi' K^+$ from BaBar, obtained by fit in bins, (b) $B^+ \to
  \gamma \psi' K^+$ and (c) $B^0 \to \gamma \psi' K^0$ from Belle.}
\label{xrad}
\end{figure}

The first evidence for $\x \to \gamma \jpsi$ by Belle was based on
$256\,\fb$ with $13.6\pm4.4$~events~\cite{ooo} and was confirmed by
BaBar on $424\,\fb$ with $23.0\pm6.4$~events~\cite{ppp}. Observation of
this channel confirmed even parity of \x. In 2009 BaBar reported
evidence of $\x \to \gamma \psi'$ based on $424\,\fb$ with $25.4\pm7.4$
signal events ($3.6\,\sigma$)~\cite{qqq} (see Fig.~\ref{xrad}, (a)). The
signal yield implied $\mathcal{B}(\x \to \gamma \psi')/\mathcal{B}(\x
\to \gamma \jpsi)=3.4\pm1.4$. However in 2010 Belle based on a larger
sample $711\,\fb$ found no evidence for $\x \to \gamma \psi'$ (see
Fig.~\ref{xrad}, (b), (c)), while $\gamma \jpsi$ mode was observed at a
rate that agrees with BaBar~\cite{rrr}. Belle set a 90\%\,CL upper limit
on the $\gamma \psi'/\gamma \jpsi$ ratio of $<2.0$.

\section{Study of $\omega \jpsi$ final state}
Three states with masses close to 3940\,\mev\ were found:
$X(3940)$~\cite{sss}, $Y(3940)$~\cite{ttt} and $Z(3930)$~\cite{uuu}, the
latter usually identified with $\chi'_{c2}$. These three states are
considered to be distinct particles, though there is no decisive
evidence for this. $Y(3940)$ mass is well above $DD$ and $DD^*$
thresholds, but the partial width of decay to hidden charm is
unexpectedly large:
$\mathcal{B}(Y\to\omega\jpsi)/\mathcal{B}(Y\to D^0D^{*0})>0.71$~\cite{vvv}.

Belle studied untagged two-photon process $\gamg\to\omega\jpsi$ with
694\,\fb\ of data, collected at $\Upsilon(4S)$, $\Upsilon(3S)$ and
$\Upsilon(5S)$ resonances. A state with $M=3915\pm4\,\mev$ and
$\Gamma=17\pm11\,\mev$ was found~\cite{www}, compatible with
$Y(3940)$. If it is so, it narrows its quantum numbers $J^{PC}$ to
$0^{\pm+}$ or $2^{\pm+}$. Measured partial width
$\Gamma_\gamg\mathcal{B}(Y\to\omega\jpsi)=61\pm19\,\ev$ (for
$0^{++}$). If $\Gamma_\gamg\sim\mathcal{O}(1\,\kev)$, a typical value
for charmonium, then $\Gamma(Y\to\omega\jpsi)\sim\mathcal{O}(1\,\mev)$,
which is very large for a hadronic inter-charmonium transition.

Though mass of \x\ is too small for decay to $\omega\jpsi$, in some
models it may decay to low-mass tail of the $\omega$ and \jpsi\ with a
rate, comparable to decay $\x\to\pi\pi\jpsi$~\cite{nnn2}. In 2005 Belle
reported an evidence for subthreshold decay $\x\to\omega\jpsi$,
consistent with the prediction~\cite{ooo}. In 2008 BaBar studied
$B$-decay $B^+\to\pi\pi\pi^0\jpsi K^+$ and in mass distribution of
$\pi\pi\pi^0\jpsi$ observed $Y(3940)$, but did not find
\x~\cite{zzz}. In 2010 BaBar remade this analysis with 433\,\fb\ and
lower requirement on $\pi\pi\pi^0$ invariant mass loosened from
0.7695\,\gev\ to 0.7400\,\gev. Both $Y(3940)$ and \x\ were observed with
masses and widths, consistent with previous measurements. BaBar also
investigated the shape of $\pi\pi\pi^0$ invariant mass distribution for
selected $\x\to\omega\jpsi$ events. They found that it favours $P$-wave
description by $1.5\,\sigma$ ($\chi^2/\mathrm{NDF}=10.17/5$ for
$S$-wave, $\chi^2/\mathrm{NDF}=3.53/5$ for $P$-wave), which indicates
$J^P=2^-$ for \x, which thus may be interpreted as $\eta'_{c2}$
charmonium state. However, possible interference between different
decays, contributing to $\pi\pi\pi^0\jpsi$ final state, was not taken
into account, and explanation of significant rate of $\x\to D\bar{D}\pi$
would be a challenge for $\eta'_{c2}$~\cite{kalash}.

\begin{figure}[htb]
\begin{center}
\includegraphics[width=0.66\textwidth,clip=true]{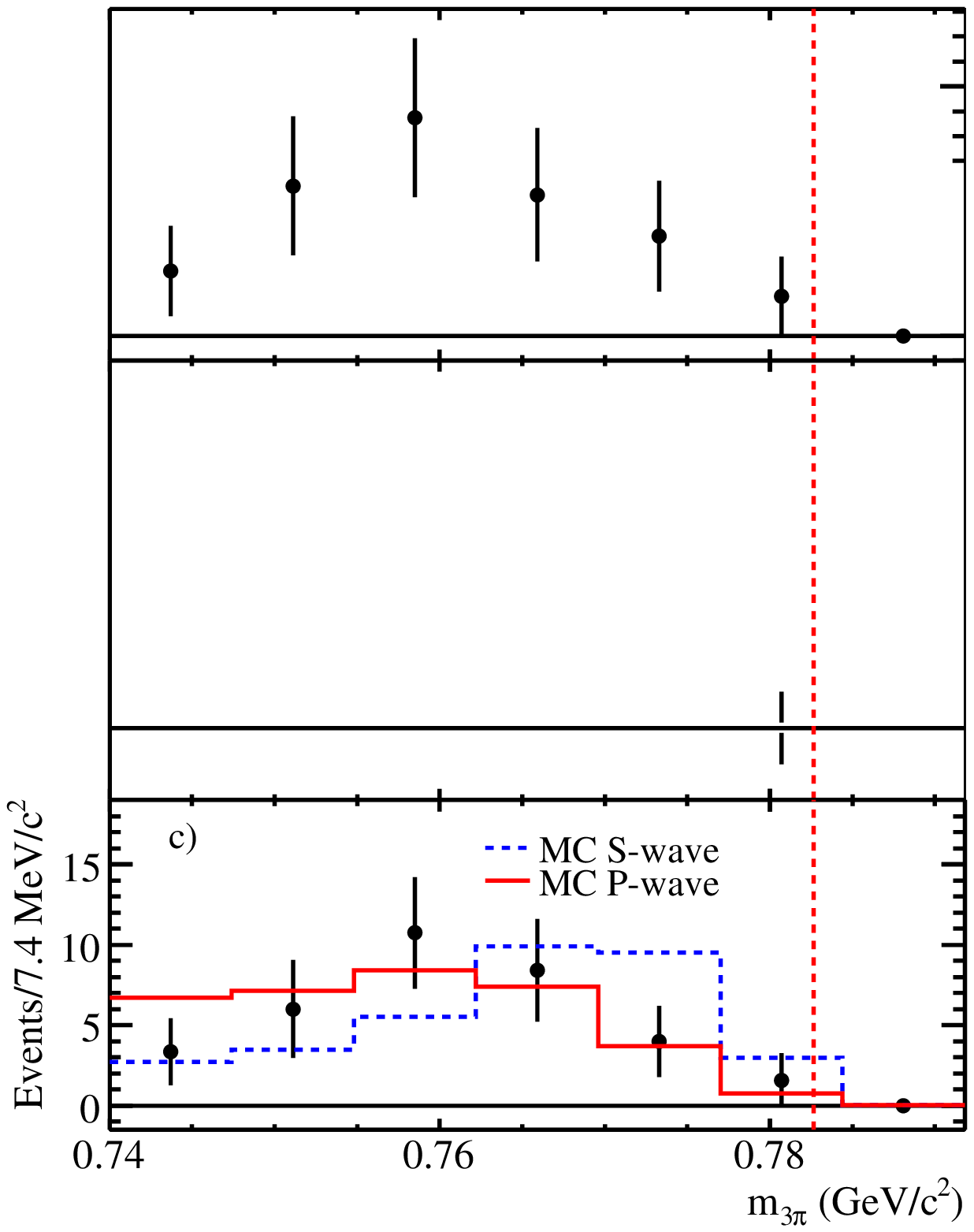}
\end{center}
\caption{The $\pi^+\pi^-\pi^0$ invariant mass distribution for $\x \to
  \pi^+\pi^-\pi^0 \jpsi$ decays from BaBar.}
\label{omega}
\end{figure}

\section{Search for charmonium production in radiative $\Upsilon$ decays}
Belle used its extensive data set, collected at $\Upsilon(1S)$
resonance, to investigate $b\bar{b}\to c\bar{c}\gamma$
transitions~\cite{222}. Calculation predicts $\sim10^{-6}$ decay rates
for lowest lying $P$-wave spin-triplet ($\chi_{cJ}$, $J=0,1,2$) and
$\sim5\times10^{-5}$ for $S$-wave spin-singlet state
$\eta_c$~\cite{333}. No prediction exists for allowed excited or
``exotic'' states, like \x. The photon detection required
$E^{lab}_\gamma>3.5\,\gev$, which corresponded to $4.8\,\gev$ mass of a
particle, produced in $\Upsilon(1S)$ radiative decay. Initial state
radiation (ISR) was removed by requirement on photon polar angle. ISR
production of $\psi'$ in $\pi^+\pi^-\jpsi$ channel was used as a
cross-check, and the cross section for this process was determined as
$20.2\pm1.1\,\mathrm{pb}$, in agreement with theoretical calculation.
One event was observed in the signal region of \x, which corresponds to
upper limit
$\mathcal{B}(\Upsilon(1S)\to\gamma\x)\times\mathcal{B}(\x\to\pi^+\pi^-\jpsi)<2.2\times10^{-6}$
at 90\%\,CL. Furthemore, no evidence for excited charmonium states
below $4.8\,\gev$ was found.

\section*{Acknowledgments}
We thank the KEKB group for excellent operation of the accelerator, the
KEK cryogenics group for efficient solenoid operations, and the KEK
computer group and the NII for valuable computing and SINET4 network
support. We acknowledge support from MEXT, JSPS and Nagoya's TLPRC
(Japan); ARC and DIISR (Australia); NSFC (China); MSMT (Czechia); DST
(India); MEST, NRF, NSDC of KISTI, and WCU (Korea); MNiSW (Poland); MES
and RFAAE (Russia); ARRS (Slovenia); SNSF (Switzerland); NSC and MOE
(Taiwan); and DOE (USA).

This work was done with partial support of Russian Federation President
grant MK-450.2010.2.

\end{document}